\documentstyle[aps,preprint]{revtex}
\begin{document}
\draft

\hsize\textwidth\columnwidth\hsize\csname @twocolumnfalse\endcsname
\title{Oscillation Frequencies for a Bose Condensate in a
Triaxial Magnetic Trap}
\author{E.~Cerboneschi$^\dagger$, R.~Mannella$^\dagger$, 
E.~Arimondo$^\dagger$
 and L.~Salasnich$^*$}
\address{
$^\dagger$Istituto Nazionale per la Fisica della Materia,
 Unit\`a di Pisa,
\\
Dipartimento di Fisica, Universit\`{a} di Pisa,
Via Buonarroti, I-56127 Pisa, Italy.
\\
$^*$Istituto Nazionale per la Fisica della Materia,
 Unit\`a di Milano,
\\
Dipartimento di Fisica, Universit\`{a} di  Milano,
Via Celoria 16, I-20133 Milano, Italy.
}
\date{\today}

\maketitle \widetext

\begin{abstract}
We investigate the dynamics  of a
Bose condensate, interacting  with either repulsive or
attractive forces,  confined in a  fully anisotropic harmonic
potential.
The (3+1)-dimensional Gross-Pitaevskii equation is integrated
numerically to derive the collective excitation frequencies, showing
a good agreement with those calculated with a
variational technique.
\end{abstract}


In Bose-Einstein condensation, an atomic sample confined within a
magnetic trap is brought into a new state of matter, described by a
single macroscopic quantum-mechanical wave function.
The condensation  process
starts from  a sample in vapour phase, and proceeds through
applications of laser cooling and evaporative
cooling \cite{anderson,bradley,davis}.
The condensate may contain
between one thousand and few millions atoms.
Around zero temperature or
when the non-condensate may be neglected, the description of the
macroscopic wave function is given by the nonlinear Gross-Pitaevskii
equation \cite{dalfovo2}.
Recently, a large attention within the Bose-Einstein
condensation community has been concentrated on the
excitation of condensate oscillations through a modulation of the
confining potential produced by the magnetic
trap \cite{mewes,jin,stamper}.
The experimental and theoretical
investigation of the condensate oscillation eigenmodes has provided
important information on the condensate.
The validity of the
theoretical treatments has been tested, the atom coupling has been
investigated in the different regimes, the processes responsible for
the damping of the eigenmodes have been determined.

In this work, we examine theoretically the condensate eigenmodes
in a magnetic trap configuration leading to a very complex
dynamics.
Most previous analyses examined magnetic traps having either
spherical or cylindrical symmetries, where the oscillation eigenmodes
are determined by the trap symmetry.
A different trap geometry, with
different elastic constants along the three orthogonal axes, has been
investigated in an experiment on sodium atoms, based on a triaxal TOP
trap\cite{phillips}.
In this letter we examine, for the first
time, the collective frequencies associated to that magnetic trap.
The
calculation of the oscillation collective frequencies requires to
solve the Gross-Pitaevskii equation (GPE) in order to determine  the
condensate spatial distribution in the ground state.
Such a  solution
makes it possible also to  investigate the modification of the
condensate distribution following a modification of the magnetic trap
geometry, for instance as a consequence of an oscillation of the trap
confining potential.
The GPE is here analyzed using the variational method based on
the Gaussian ansatz for the shape of the macroscopic
wave function of the condensate, applied to the study of the
nonlinear Schr\"odinger equation in nonlinear optical wave
propagation \cite{variational} and for the investigation of the
Bose-Einstein condensate in cylindrical \cite{zoller} and spherical
\cite{sala} symmetry traps. 
This variational approach is here extended to
the triaxial magnetic trap.
 We derive the low-energy excitation
spectrum of  the collective motion for both positive and negative
scattering lengths, as a function of the atom-atom interaction 
strength. 
Our analysis is valid for the whole range of coupling
strengths, and, in the limit of large coupling, we recover the 
predictions of the hydrodynamic approach in Thomas-Fermi
approximation \cite{dalfovo2}. 
We have also solved numerically the
(3+1)-dimensional GPE, and derived the condensate mode frequencies 
examining the oscillation frequencies excited in the temporal
evolution of the condensate wavefunction. 
The agreement between the
collective mode frequencies calculated with the approximate
variational technique and those arising from the exact numerical
integration is within a few percent.

The macroscopic
condensate wave function $\psi$, describing an average number $N$ of
interacting bosons trapped in an external potential $V$ at zero
temperature, obeys the time-dependent GPE
\begin{equation}
i \hbar \frac{\partial}{\partial t} \psi(\vec r, t) =
\left( - \frac{\hbar^{2}\nabla^{2}}{2 m} + V(\vec r) + g |\psi
(\vec r,t)| ^{2}\right) \psi (\vec r,t),
\label{gross}
\end{equation}
where $m$ is the mass of the atom and the coupling constant $g$ is
given by  $g=4\pi\hbar^2a/m$, with $a$ the $s$-wave scattering length.
The time-dependent GPE  can  be derived by minimizing the action
$S=\int{\cal L}\; d^3\vec{r}dt$
related to the Lagrangian density
\begin{equation}
{\cal L}=
i\hbar \psi^{\ast}\frac{\partial\psi}{\partial t}
-\frac{\hbar^2}{2m}\vec{\nabla}\psi^{\ast}\cdot\vec{\nabla}\psi
-V\left(\vec{r}\right)\left|\psi\right|^2
-\frac{g}{2}\left|\psi\right|^4.
\end{equation}
We consider a triaxially asymmetric harmonic trapping potential of
the form
\begin{equation}
V\left(\vec{r}\right)=
\frac{1}{2}m\omega_0^2\left(\lambda_1^2 x^2+
\lambda_2^2 y^2+\lambda_3^2 z^2\right),
\end{equation}
where the adimensional constants $\lambda_i^{2}$ ($i=1,2,3$) are 
proportional to the spring constants of the potential along the three
axes.

Minimizing the action $S$ within a Gaussian trial function
makes it possible to obtain, for the condensate
wave function, approximate solutions  which can be derived, to a large
extent, analytically  \cite{variational,zoller,sala}.
In the
limit of weak interatomic coupling, i.e., small boson number
and small scattering length, the choice of a Gaussian shape for the
condensate is well justified because it  describes the
exact solution of the linear Schr\"odinger equation with harmonic
potential. 
For the description of the collective dynamics of a
Bose-Einstein condensate trapped in a spherically or cylindrically
symmetric potential, it has already been shown that the
variational technique based on Gaussian trial functions leads
to reliable results even in the large condensate number
limit~\cite{zoller}.
We take a trial function of the form
\begin{eqnarray}
\psi\left(\vec{r},t\right) & = & N^{1/2}
\Big( {1\over \pi^3 \tilde{\sigma}_1^2(t) \tilde{\sigma}_2^2(t)
\tilde{\sigma}_3^2(t)} \Big)^{1/4} \times \nonumber \\
& &\prod_{i=1,2,3}
\exp\left\{-\frac{x_i^2}{2\tilde{\sigma}_i^2(t)}
+i \beta_i(t) x_i^2 \right\},
\label{ansatz}
\end{eqnarray}
with $(x_1,x_2,x_3)\equiv(x,y,z)$,
where $\tilde{\sigma}_i$ and $\beta_i$ are
time-dependent variational parameters and the wave function
is normalized to the number $N$ of bosons.
Please note that,
in order to describe the time evolution of the variational
function, the phase factor $i \beta_i(t) x_i^2$ is needed
\cite{variational}.

>From the Euler-Lagrange equations, the following equations of
motions for the variational parameters are derived:
\begin{mathletters}
\begin{eqnarray}
\ddot{\tilde{\sigma}}_i(t)+\lambda_i^2\omega_0^2\tilde{\sigma}_i(t)
&=&
\frac{\hbar^2}{m^2\tilde{\sigma}_i^3(t)}
+ \nonumber \\
& &\sqrt{\frac{2}{\pi}}
\frac{a\hbar^2N }
{m^2\tilde{\sigma}_i(t)\tilde{\sigma}_1(t)\tilde{\sigma}_2(t)
\tilde{\sigma}_3(t)}
\label{stilde} , \\
\beta_i(t)
&=&
-\frac{m\dot{\tilde{\sigma}}_i(t)}{2\hbar^2\tilde{\sigma}_i(t)},
\label{beta}
\end{eqnarray}
\end{mathletters}
with $i=1,2,3$.
>From Eq.~(\ref{beta}), we observe that the time dependence of
 $\beta_i$ is determined by that of
$\tilde{\sigma}_i$.
Introducing the adimensional parameters $\tau=\omega_0t$ and
$\sigma_i=\tilde{\sigma}_i/a_0$, where $a_0=(\hbar/m\omega_0)^{1/2}$
is the harmonic oscillator length,
Eq.~(\ref{stilde}) becomes
\begin{equation}
\frac{d^2}{d\tau^2}\sigma_i+\lambda_i^2\sigma_i=
\frac{1}{\sigma_i^3}
+\frac{\tilde{g}}{\sigma_i\sigma_1\sigma_2\sigma_3}.
\label{sigma}
\end{equation}
Here, we have defined the coupling strength
$\tilde{g}=(2/\pi)^{1/2}Na/a_0$,  proportional to the
condensate number $N$ and the scattering length $a$, which determines
the atom-atom interaction compared to the bare trapping potential.
Equations (\ref{sigma}), with $i=1,2,3$,  represent a set of
three nonlinearly coupled ordinary differential equations describing
the  time evolution of the widths of the condensate along each
direction \cite{note}: formally, they correspond to the classical
equations of motion for a particle with coordinates $\sigma_i$ and
total energy $E=T+U$, with
\begin{mathletters}
\begin{eqnarray}
T
&=&
\frac{1}{2}\left(\dot{\sigma}_1^2+\dot{\sigma}_2^2+
\dot{\sigma}_3^2\right),
\\
U
&=&
\frac{1}{2}\left(\lambda_1^2\sigma_1^2+\lambda_2^2\sigma_2^2
+\lambda_3^2\sigma_3^2\right) \nonumber \\
& & +\frac{1}{2}\left(\frac{1}{\sigma_1^2}+\frac{1}{\sigma_2^2}
+\frac{1}{\sigma_3^2}\right)
+\tilde{g}\frac{1}{\sigma_1\sigma_2\sigma_3},
\label{poten}
\end{eqnarray}
\end{mathletters}
where the dots indicate the derivative with respect to $\tau$.
$T$ plays the role of kinetic
energy and the three terms in $U$, arising from the harmonic trapping
potential, the kinetic pressure, and the atom-atom interaction,
respectively, represent an effective potential energy.

The frequencies of the
low energy excitations of the condensate correspond to
the small oscillations of the variables $\sigma$'s around the 
equilibrium point, given by the minimum of the effective potential
energy $U$ in Eq.~(\ref{poten}).
First, we find the equilibrium point solving the equations
\begin{equation}
\lambda_i^2\sigma_i^4
-\tilde{g}\frac{\sigma_i}{\sigma_j\sigma_k}
=1,
\label{minimum}
\end{equation}
with $i,j,k$ ranging from $1$ to $3$ and different from each other.

Then, after a second-order Taylor expansion of $U$ around the
minimum, we linearize Eqs.~(\ref{sigma}) around the stationary
point. The calculation of the normal mode frequencies for the motion 
of the condensate is reduced to an eigenvalue problem for the matrix
$\Lambda$, given by
\begin{equation}
\Lambda_{ij}=
\left.\frac{\partial^2U}{\partial\sigma_i\partial\sigma_j}
\right|_{\vec{\sigma}=\vec{\sigma}^0},
\label{eigenvalue}
\end{equation}
where $\vec{\sigma}^0=(\sigma_1^0,\sigma_2^0,\sigma_3^0)$ stands for
the solution of Eq.~(\ref{minimum}).
In the Thomas-Fermi limit, i.e., for $\tilde{g}\gg 1$,
 the right-hand
side   in Eq.~(\ref{minimum}), related to the kinetic pressure, can be
neglected and an analytic expression for the
equilibrium point is obtained.
In this case, the eigenfrequencies $\omega$
for the collective motion, in units of $\omega_0$, are given by the
solutions of the equation
\begin{eqnarray}
0 &=& \omega^6
-3\left(\lambda_1^2+\lambda_2^2+\lambda_3^2\right)\omega^4
 \nonumber \\
& &+8\left(\lambda_1^2\lambda_2^2+\lambda_1^2\lambda_3^2
+\lambda_2^2\lambda_3^2\right)\omega^2
-20\lambda_1^2\lambda_2^2\lambda_3^2.
\end{eqnarray}
The same equation, obtained within a hydrodynamic approach, is
reported in Ref.~\cite{dalfovo2}.
In the opposite limit,
for $\tilde{g}= 0$, we recover
the well-known result $\omega_i=2\lambda_i$, with $i=1,2,3$, in units
of $\omega_0$, for  the normal modes in the non-interacting case.

\begin{figure}
\caption[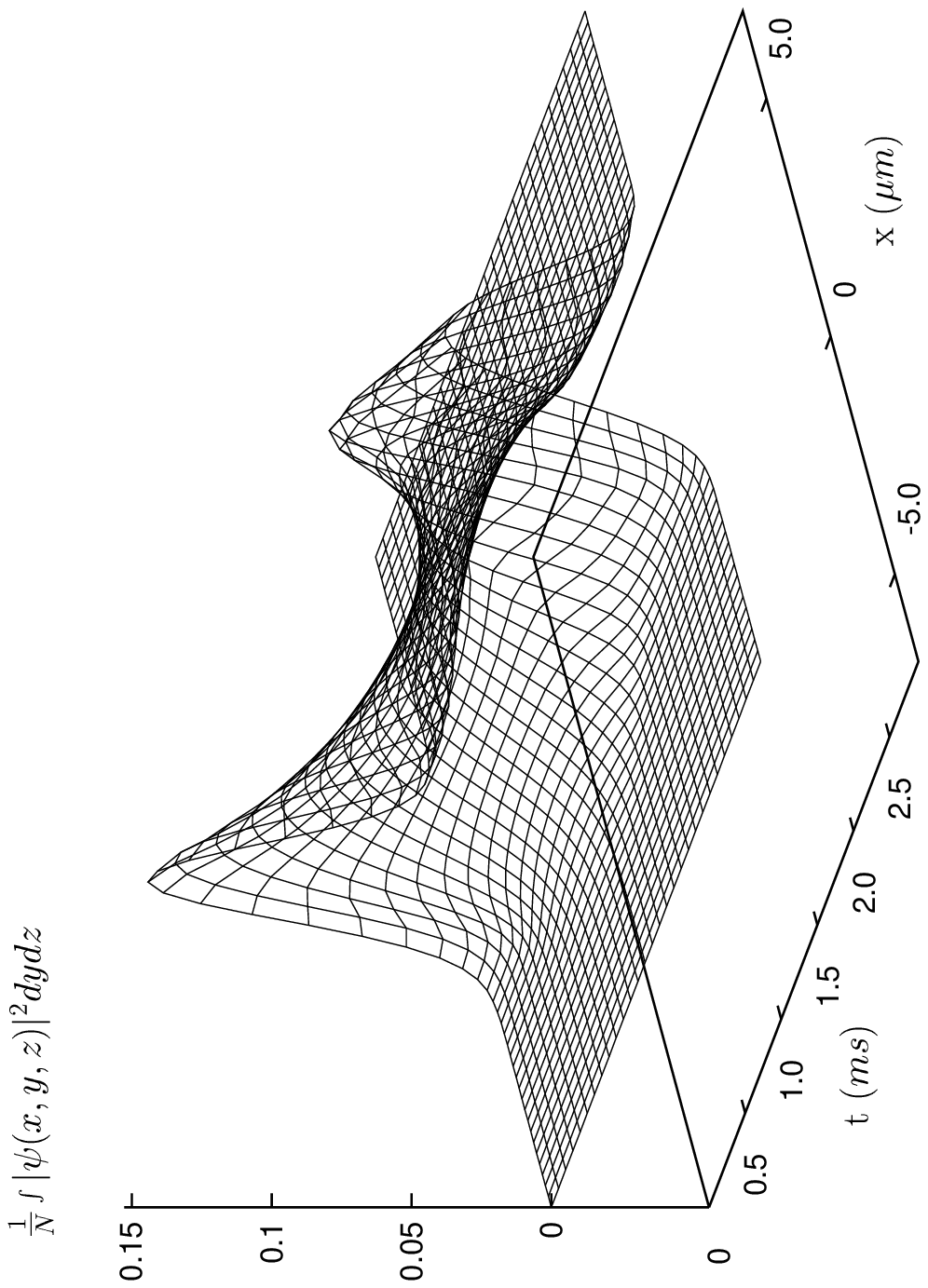]{A typical result of the numerical integration of the
GPE: plot of $N^{-1}\int |\psi(\vec r,t)|^{2} \;dy \;dz$
vs $t$ and $x$. We used the Na parameters for the
integration (see text), and $\tilde g=3.335$.}
\label{psi}
\end{figure}

The numerical integration of the time dependent GPE, in
Eq.~(\ref{gross}), was done using a modified split operator technique,
adapted to the integration of a Schr\"odinger equation.
We wrote Eq.~(\ref{gross})
in the form
\begin{equation}
i \hbar \frac{\partial}{\partial t} \psi(\vec r, t) =
\left( H_{x}(\vec r,t) + H_{y}(\vec r,t) + H_{z}(\vec r,t)\right) \psi
(\vec r,t),
\label{gpenum}
\end{equation}
where
\begin{equation}
H_{i} (\vec r,t) \equiv - \frac{\hbar^{2}}{2m}\frac{ \partial^{2}}
{\partial
x_{i}^{2}} + V(x_{i}) + \frac{1}{3} g  |\psi
(\vec r,t)| ^{2}.
\end{equation}
The idea is to split the full Hamiltonian in three sub-Hamiltonians, 
so that at each time we have to write the Laplacian with respect to 
one coordinate only, leading to the solution of a
tridiagonal system, and to huge savings in computer memory.
The splitting is carried out so that the
commutators are exact up to the order $\delta ^{2}$ included.
Equation (\ref{gpenum}) was integrated using the scheme
($\delta$ is the integration time step, and $A_{i}(t)\equiv i \delta
H_{i}(\vec r,t)/\hbar$)
\begin{eqnarray}
\psi (\vec r, t + \delta)
&=&\frac{1}{1+A_{y}(t)/2}\left(1-A_{x}(t)/2\right) \times \nonumber \\
& &\frac{1}{1+A_{z}(t)/2}\left(1-A_{z}(t)/2\right) \times \nonumber \\
& &\frac{1}{1+A_{x}(t)/2}\left(1-A_{y}(t)/2\right) \psi (\vec r ,t).
\end{eqnarray}
There is
obviously a problem with the nonlinear term $g |\psi (\vec r,
t)|^{2}$, because we should really use a $\psi$ somehow averaged over
the time step $\delta$, not a $\psi$ evaluated at the beginning of the
time step.
To circumvent this problem, we used a sort of predictor
corrector step.
Each integration step is really done in two times:
going from the time $t$ to the time $t+\delta$, the first time we
used $\psi (\vec r, t)$ in the nonlinear term, obtaining a
``predicted'' $\tilde \psi (\vec r, t+\delta)$; we then repeated the
integration step, starting again from $\psi (\vec r, t)$, but using
$\frac{1}{2} \left (\psi (\vec r, t) + \tilde \psi (\vec r,
t+\delta) \right )$ in the nonlinear term.
In Fig.~\ref{psi} we show a typical time evolution obtained from the
integration of the GPE. We plotted here $N^{-1}\int |\psi(\vec
r,t)|^{2} \;dy \;dz$
as function of $t$ and $x$. 
The initial wave function we started from
has the form of
Eq.~(\ref{ansatz}), with $\sigma_{i}=1$ and $\beta_{i}=0$ for
$i=1,2,3$, i.e., fairly far from equilibrium.

\begin{figure}
\caption[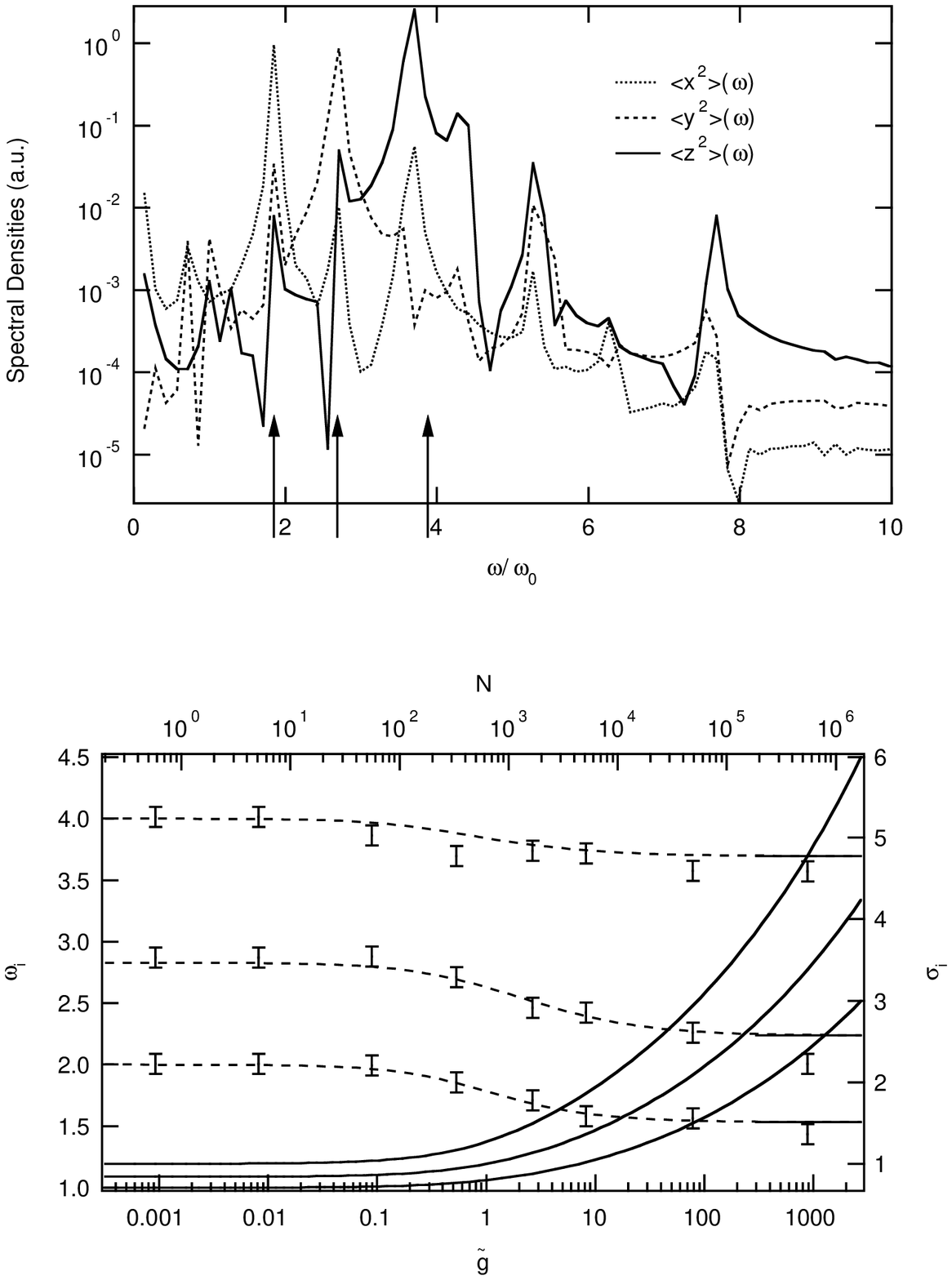]{Top: Typical spectral densities from the
numerical integration of the GPE: we show the Fourier transform of the
time evolution of the condensate widths along the three axes, for the 
Na case and
$\tilde g = 0.664$. 
The arrows show the location of the eigenmode
frequencies expected from the variational analysis.
Bottom: Results for the variational
calculation of the scaled oscillation frequencies $\omega_i$, in units
of $\omega_0$ (dashed lines), and scaled widths $\sigma_i$, in units
of $a_0$ ( continuous lines), as a function
of ${\tilde g}$ (lower scale) and $N$ (upper scale). 
The horizontal
segments on the right represent the hydrodynamic limits. 
The markers
indicate the oscillation frequencies derived from numerical
integration of the GPE. 
Triaxial TOP parameters as in \cite{phillips}
and scattering length $a=2.75$~nm.}
\label{pos}
\end{figure}

The Gaussian wave function is not
an exact steady-state solution for the
GPE, as also stated in \cite{singh}.
The small
deviations from the actual stationary solution excite small-amplitude
oscillations in the condensate distribution.
We have
spectrally analyzed the time dependence of the condensate
widths along the three spatial directions, in
order to derive the oscillation frequencies.
This analysis has
been done  for  different values of $\tilde{g}$, for both positive
and negative scattering lengths.  A typical spectral
density is shown in Fig.~\ref{pos} (top).

\begin{figure}
\caption[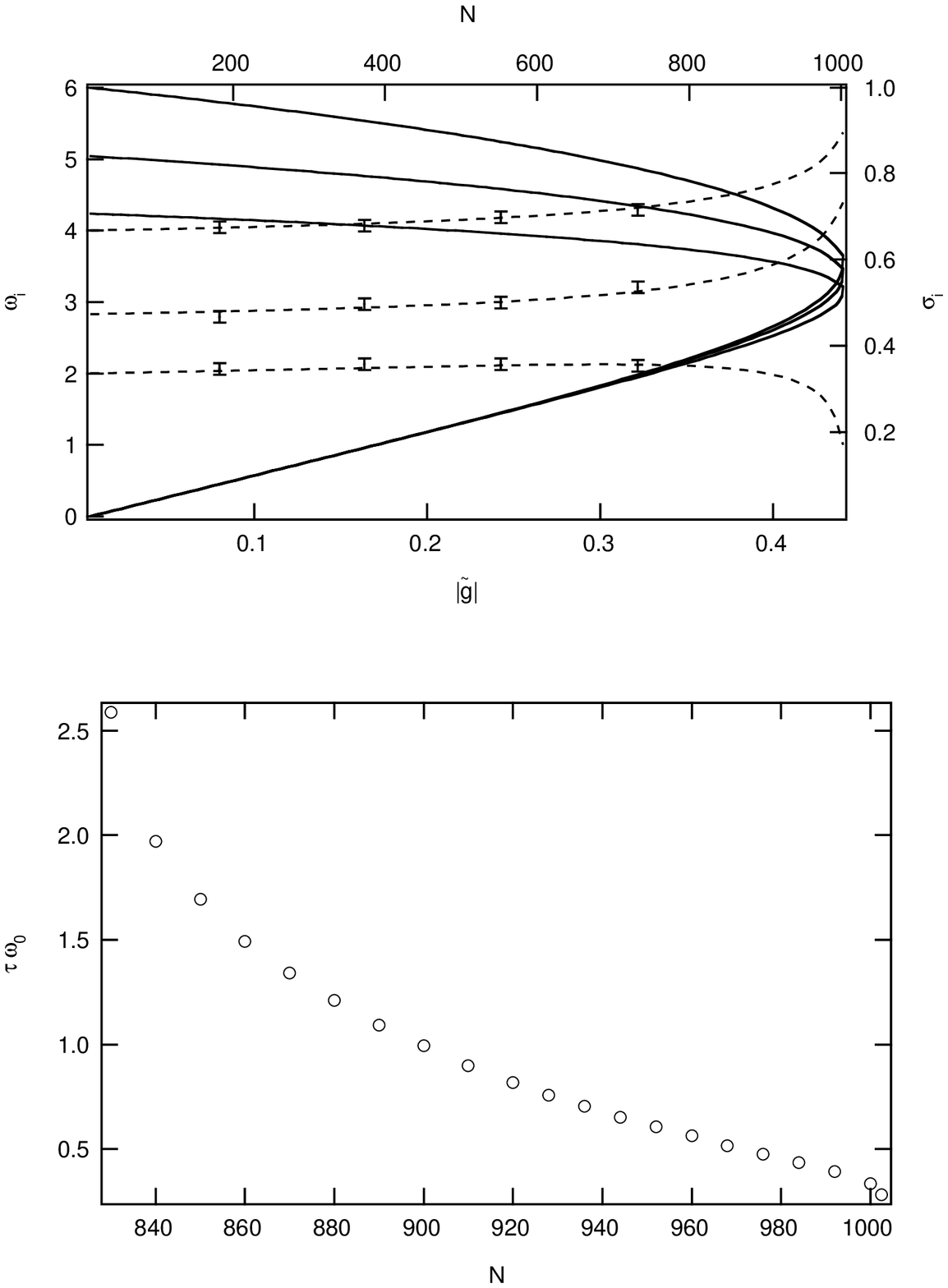]{Top: As in Fig.~\protect{\ref{pos}}, except 
scattering length $a=-1.4$~nm. 
Bottom: time before we observe the
condensate collapse, as a function of $N$, for the same parameters of
the top figure. 
The
initial wave function has the form of Eq.~(\protect{\ref{ansatz}}),
with the $\sigma_{i}$ equal to the variational ones, and the
$\beta_{i}=0$.}
\label{neg}
\end{figure}

The general solution of Eq.~(\ref{minimum})
for the stationary state and of Eq.~(\ref{eigenvalue}), the eigenvalue
problem for the collective mode frequencies, was obtained
numerically.
This solution is shown in Fig.~\ref{pos} (bottom), where
the scaled widths $\sigma_i$ of
the stationary condensate wave-function, solid lines,
and the normal
mode frequencies $\omega_i$, dashed lines, are shown
as functions of the coupling strength $\tilde{g}$ in case of a 
positive scattering length $a$.
Figure \ref{neg} (top) shows  the same
quantities for a negative scattering length $a$. 
On the top axes of the
graphs in Figs.~\ref{pos} (bottom) and \ref{neg} (top), the number $N$ 
of atoms in
the condensate is reported, for scattering lengths and
trapping potential parameters fixed to the values listed below. 
The
markers in Figs.~\ref{pos} and \ref{neg} represent the mode
frequencies calculated from the numerical solution of the GPE.
The results in Fig.~\ref{pos} are obtained for parameters
corresponding to the triaxially anisotropic trap used in
the experiment in Ref.~\cite{phillips}, performed at
NIST on Na atoms, e.g.,
$\omega_0=2\pi\times 234$~s$^{-1}$, $\lambda_1=1$,
$\lambda_2=\sqrt{2}$ and $\lambda_3=2$.
The Na scattering length $a=2.75$~nm is used in the calculation.
The equilibrium solutions for the condensate widths $\sigma_i$
are  increasing functions of the condensate number $N$ and,
asymptotically,  scale as $N^{1/5}$.
The mode frequencies decrease with $N$
monotonically from the non-interacting values to the asymptotic
values, indicated in the figure by the
solid horizontal segments.

Figure~\ref{neg} (top) shows the solution for a negative scattering 
length $a=-1.4$~nm, appropriate for $^7$Li atoms.
The parameters characterizing the trapping potential are the same as
in Fig.~\ref{pos}.
For small condensate numbers
we recover again the non-interacting limit.
With increasing $N$, the  atom-atom interaction strongly
modifies the dynamics of the condensate, in a different way with
respect to the case of positive scattering length.
Notice the different
horizontal scales in Figs.~\ref{pos} and~\ref{neg}.
It is worth noting that the variational formalism, being applicable
for any condensate number $N$, makes it possible to
investigate the case of negative scattering lengths.
In this case, the large $N$ limit of the Thomas-Fermi approximation
and the hydrodynamic approach do not apply, because
the number of condensed atoms cannot exceed a critical 
value~\cite{dalfovo2}.
 For negative coupling constant $\tilde{g}$,
Eqs.~(\ref{minimum}) present either two solutions  for the stationary
width $\sigma_i$ of the condensate along each  direction, or no
solution, depending on the  absolute value of $\tilde{g}$.
This is shown in Fig.~\ref{neg}, where, for each
$\sigma_i$, a two-branched solution is displayed (solid lines).
The upper and lower branches of the $\sigma_i$ curves correspond to
the  stable and the unstable solution, respectively.
As functions of $|\tilde{g}|$, the
stable stationary widths shrink, while the
unstable ones, originating from zero as a limit for
$|\tilde{g}|\rightarrow 0$, increase.
The two branches of each $\sigma_i$ solution merge at a  critical
value of the  coupling strength.
Beyond this value,  no solution to Eqs.~(\ref{minimum}) exists.
The dashed lines in Fig.~\ref{neg} show the frequencies of the 
collective motion
around the stable solution.
At the critical point, the two highest frequencies diverge and the
lowest one falls to zero.
Similar results have been obtained in Ref.~\cite{zoller} for
cylindrically symmetric condensates. This theoretical results agree
well with the numerical integration of the GPE.

However, in the case of negative scattering lengths the collapse of 
the condensate implies that the GPE can be numerically integrated only
until the collapse takes place. 
As the number of particles in the
condensate increases and approaches the critical value, the lifetime
of the condensate
decreases and, eventually, becomes so short that we can no longer
observe any oscillatory dynamics. 
For the Li parameters, the collapse
lifetime is shown in Fig.~\ref{neg} as function of $N$, for $N$
approaching the critical value.

In conclusion, we have performed an analysis of the dynamics of a
Bose-Einstein condensate trapped in a triaxially
asymmetric potential.
Condensation was
achieved at NIST in a fully asymmetric TOP trap \cite{phillips}, while
a specific study of the condensate dynamics in such a geometry
was lacking.
At present, no experimental data concerning the collective
motion of the condensate are available from the
NIST experiment.
Our predictions might be confirmed by
forthcoming experimental results.
Within the different point of view started by
Dalfovo {\it et al.}\cite{dalfovo1}, the study of the
nonlinear coupling between the condensate modes provides a stronger
test of the condensate characteristics. 
Moreover, nonlinear processes
could be useful to perform nonlinear spectroscopy or nonlinear optics
in the condensate. 
As well known in the framework of nonlinear
dynamics, more complex scenarios are produced
when the number of degrees of freedom is increased, as in the
(3+1)-dimensional evolution in a triaxial magnetic
trap.

This project was supported by the Italian INFM through a PRA project 
on Bose-Einstein
Condensation and by the CNR through a Progetto Integrato.
The authors wish to thank L.~Reatto for a careful reading of the
manuscript.

\end{document}